\documentstyle[prl,twocolumn,aps]{revtex}
\input epsf
\begin{document}
\draft

\title{Slow light in three-level cold atoms: a numerical analysis}

\author{E. Cerboneschi$^{(1)}$, F. Renzoni $^{(2)}$
,   and E. Arimondo$^{(1)}$}
\address{\ $^{(1)}$ INFM
and Dipartimento di Fisica, Universit\`a di Pisa, Via F. Buonarroti
2, I-56127 Pisa,
Italia}

\address{\ $^{(2)}$ Laboratoire
Kastler-Brossel, Departement de Physique de l'Ecole Normale Sup\'erieure,
24 rue Lhomond, 75231 Paris Cedex 05 France}

\date{\today{}}
\maketitle

\begin{abstract}
We investigate theoretically the slow group velocity of a
pulse probe
laser propagating through a cold sample and interacting with atoms in a
three-level
$\Lambda$ configuration having losses towards external states. The EIT
phenomenon produces very small group velocities for the probe pulse in
presence of a strong coupling field even in presence of the population
losses, as in an open three-level system.  The group velocity and the
transmission of the pulses are examined
numerically as functions of several parameters, the adiabatic
transfer, the loss rate, the modification of the atomic velocity
produced within the cold sample. The conditions for a more
efficient pulse transmission through the cold atomic sample are specified.

\end{abstract}

\section{Introduction}
In a  a three level $\Lambda$ system, with a central  excited level connected
by electric dipole transitions to the two ground ones,
electromagnetically-induced transparency (EIT) and  reduction in group velocity
are based on the low-frequency coherence created by the laser radiation
\cite{harris_rev,scullybook}. One optical transition is driven by a strong
laser,
denoted as coupling laser, and the second transition is driven by a
weak laser, the probe one. The probe laser absorption and dispersion are
determined by the modifications in the populations and in the coherences
produced
by the coupling laser, and also by the low-frequency coherence between the
ground states created by the simultaneous application of the coupling
and probe lasers. For the three-level atomic $\Lambda$  configuration, the
absorption is decreased by the coherent trapping of the
population in the ground state \cite{arirev}. The probe light
group velocity is greatly reduced by the very steep dispersion associated
with the coherent population trapping narrow resonance.  Remarkable
results for the reduction of the group velocity in three-level systems
were recently reported in cold samples\cite{lau99,inouye00,liu01} as well
as in samples with
thermal velocity distributions \cite{scully99,budker99,phillips01}.

Often the real atomic transitions of quantum optics
phenomena  do not correspond to the simple theoretical
models, and several
multilevel atomic structures can be described as
open systems, i.e., systems where  population, whence coherent
population trapping, is lost because of
decay into sink levels not excited by the lasers.
Even if the open
features of a multilevel system may be
eliminated through the application of a repumping radiation to deplete the sink
state \cite{renzoni1}, they play an important role in most experiments.
In this work we plan to investigate the propagation of slow light in
an open atomic system for the specific
 conditions of the cold atom experiment by Lau et al \cite{lau99}. We
analyze a sample of cold sodium atoms laser excited in an open  $\Lambda$
scheme starting from two different
hyperfine levels and with losses from the excited state towards
another ground state. The experiment of \cite{lau99} was performed
with both a cold sodium atomic sample and a sodium condensate. However
the atomic interactions are not included in our analysis \cite{morigi}.

The present work addresses several issues related to the slow
light propagation.  Our main aim is to determine the role of the open
three-level
system losses on the  group velocity, probe pulse amplitude transmission
and transmission
bandwidth. In an open three-level system, any excited state occupation leads
to a loss of the atomic population towards atomic states not excited
by the laser light, whence to a modification of the slow light
propagation.  Our  numerical simulations  determine the laser parameters
more  appropriate to realize  both efficient pulse
slowing and transmission.  We have specifically investigated if
the STIRAP configuration\cite{bergmann,fleischh99} could be used to reduce
the role of the
excited state occupation in an open three-level system.  In STIRAP
the counterintuitive coupling/probe pulse sequence with detuned lasers
produces a very efficient coherent-state atomic preparation from the
initial ground state to the final ground state, and decreases
the role of the excited state
occupation with losses towards external levels.  We have verified
that, even if the counterintuitive pulse sequence enhances the probe pulse
transmission,  the EIT conditions for slow light do not correspond
precisely to those for the STIRAP process.
We investigate also the influence of the adiabatic transfer conditions
\cite{fleischh96} on the pulse transmission: a larger probe pulse
amplitude transmission  is obtained for a larger temporal width of the
probe pulse, because that width modifies the adiabaticity
condition.

The transmission of a slow probe pulse depends  on the decay rate of the
coherence
between the lower levels of the  $\Lambda$ system. An efficient pulse
propagation,
{\it i.e.} a small  absorption coefficient, is obtained by reducing the
coherence decay
rate. In a sample of room temperature atoms, collisions and transit  time are
the main contributions to that decay rate. We examine the influence of a
coherence relaxation
rate on the pulse transmission and on the velocity reduction. In a cold
atom  sample,
the previous contributions to the coherence decay rate are greatly
decreased.
However in a cold sample
with long interaction times the kinetic energy motion of the atoms
represents another important  contribution to the coherence
decay rate.  The
forces by the laser on the atoms modify the atomic momentum and
whence the parameters of the laser atom interaction.  We
investigate the pulse propagation taking into account the atomic
momentum modifications occurring in the laser interaction. The
importance of light forces on atoms in the case of slow light
propagation has been recently investigated \cite{harris00} for the
configuration where a slowly propagating laser pulse produces a force
acting on a second atomic species contained within the propagation medium.
In the present work the light forces act on the same atoms producing
the slow light process. Thus we investigate the limitations imposed on
the slow light propagation by the light forces acting on  the momentum
of the atoms producing the light slowing.

The present analysis makes use of the  time/space dependent
numerical solution  the Optical Bloch Equations (OBE)
of an open three-level system.
Section II introduces the three-level system and the coupling/probe
laser propagation.  Section III
presents the numerical results for the propagation of a laser pulse through
a cold
sample of open three-level atoms in the presence of a coupling laser.
 Section IV investigates the role of the
atomic momentum. A conclusion completes our work in
Section V.

\section{Three-level system and laser propagation}
\subsection{Atoms}

\begin{figure}
\centering\begin{center}\mbox{\epsfxsize 3.2 in
\epsfbox{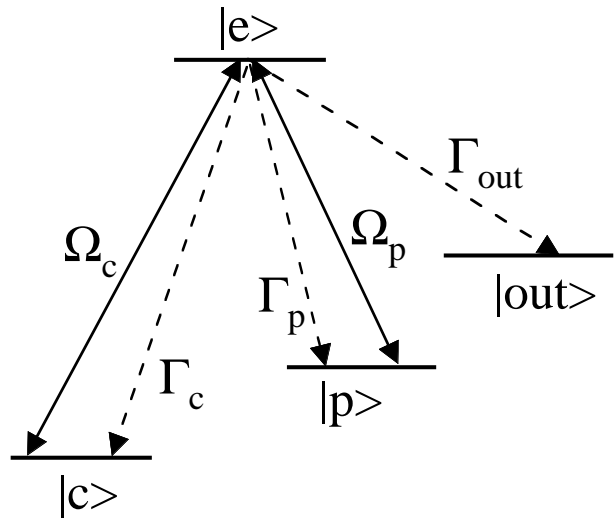}}
\end{center}
\caption{Schematic representation of an open three-level atomic system
interacting with two laser
fields in the $\Lambda$ configuration, with the decay rates $\Gamma_{\rm
c}$, $\Gamma_{\rm p}$ and $\Gamma_{\rm out}$ towards lower states.}
\label{scheme}
\end{figure}

We consider a cold sample of open three-level atoms interacting with
two laser fields in the $\Lambda$ configuration.
The interaction scheme is shown in Fig. \ref{scheme}. The ground (or
metastable) states $| \rm c\rangle$ and $| \rm p \rangle$, with energies
respectively $E_{\rm c}$ and $E_{\rm p}$, are excited to a common
state $|{\rm e}\rangle$, with energy $E_{\rm e}$, by two
laser fields
of frequencies $\omega_{\rm c}$, $\omega_{\rm p}$, electric field amplitudes
${\cal E}_{\rm c}$, ${\cal E}_{\rm p}$, and wavevectors $k_{\alpha} =
\omega_{\alpha}/c$,
with $(\alpha = {\rm c,p})$, $c$ being light speed in vacuum.
As standard in laser without inversion and similar
processes\cite{harris_rev,scullybook}, the laser fields are indicated
as  coupling ($\rm c$) and probe ($\rm p$) lasers.
The laser detunings are denoted by
$\delta_{\alpha}=\omega_{\alpha}-\omega_{\rm e\alpha}$ and
the Raman detuning from the two-photon resonance is
$\delta_R= \delta_{\rm c}-\delta_{\rm p}$.
The Rabi frequencies are given  by
\begin{equation}
\Omega_{\alpha}=\frac{{\cal D}_{{\rm e} \alpha}{\cal E}_{\alpha}}{\hbar},
\label{rabi}
\end{equation}
with ${\cal D}_{{\rm e}\alpha}$ the atomic dipole moment, supposed to be
real for the sake of simplicity.
The evolution of the atomic density matrix $\rho$ is described by the OBE
for an open system \cite{renzoni00}.
We assume the excited state relaxation, denoted by $A$,  due to spontaneous
emission,
as for a dilute atomic sample. In the open system the $A$ rate is composed
by the rates $\Gamma_{\rm c}$
and $\Gamma_{\rm p}$ for the decays  into the ground states, and by
the  decay $\Gamma_{\rm out}$
into a sink state $|out\rangle$, not excited by the lasers
\begin{equation}
A =
\Gamma_{\rm c}+\Gamma_{\rm p}+\Gamma_{\rm out}.
\label{open}
\end{equation} Thus the total decay rate
of the excited state population, denoted by $\Gamma_{\text e}$, is equal
to $A$, and the decay rate of the optical coherences,
$\rho_{\rm ep}$ and    $\rho_{\rm ec}$,  is equal to $A/2$. For a given open
system the application of a repumping laser,
as performed in \cite{renzoni1}, eliminates the decay channel out of the
system.  Thus we may  compare the slow light
propagation in an open system characterized by the above relaxation rates,
to the propagation in the closed system produced by the repumping laser.
That closed system is characterized by the following  population decay for the
excited state:
\begin{equation}
  \Gamma^{'}_{\text e} = \Gamma_{\text  c}+\Gamma_{\text p},
\label{opendecay}
\end{equation} while in the presence of repumping the decay rate of
the optical coherences remains $A/2$. In our analysis we will introduce
also a
decoherence rate $\gamma_{\rm cp}$ for
the ground state coherence $\rho_{\rm cp}$.

The response of the three-level system is
determined by the presence
of a noncoupled state, {\it i.e.} a state nonabsorbing the laser radiation,
 given by
\begin{equation}
|{\rm NC}(t)\rangle = {1\over \Omega}
\left( \Omega_{\rm p} |c\rangle
-\Omega_{\rm c} e^{i\left( \omega_{\rm p}-\omega_{\rm c} \right)t}|p\rangle
 \right) e^{-i\frac{E_{\rm c}t}{\hbar}}.
\label{nonabsor}
\end{equation}
where $\Omega$ is defined by
\begin{equation}
    \Omega =\sqrt{|\Omega_{\rm p}|^2+|\Omega_{\rm c}|^2}.
\end{equation}

The decoherence rate $\gamma_{\rm cp}$ couples $|{\rm NC}(t)\rangle$
to its orthogonal state $|{\rm C}(t)\rangle$.

In both closed and open three-level systems an optical pumping preparation
of the noncoupled
state, taking place within few excited state lifetimes, requires a significant
occupation of  excited state population\cite{arirev}.  While in a closed
system
all the atomic population falls into the  noncoupled state without loss of
atomic population, for  an open system  a  decay out
of the three-level system takes place. For an open system,   the occupation
of the ground states $|c\rangle$ and $|p\rangle$ at long interaction
times is different from zero only at small Raman detunings, where
coherent population trapping is efficient.  The analysis of
Ref.\cite{renzoni00} indicates  that at long interaction times
the slope of ${\rm Re}(\tilde{\rho}_{\rm ep})$ at $\delta_{\text R} \ne
0$ reaches a
constant nonzero value. In effect  around $\delta_{\rm R}=0$ the population is
trapped in the nonabsorbing state up to an interaction time $\Theta$ dependent
on $\delta_{\rm R}$\cite{renzoni00,renzoni98}. A constant, and slow, group
velocity is
realized because the slope of the dispersion around  $\delta_{\rm R}=0$
 does not change significantly for an increasing interaction time
$\Theta$,
even if the interval of sharp variation of the dispersion decreases with
$\Theta$.
However,  the  loss rate out of the three-level system increases
with the Raman
detuning, and  the transparency window narrows for increasing interaction time.

\subsection{Lasers}
To describe  the propagation of laser pulses through a cold sample of open
three-level
atoms, the electric field amplitudes, whence the Rabi frequencies,    are
assumed dependent on
the time $t$ and the coordinate $z$ along the propagation direction.
For slowly varying electric field amplitudes, the Maxwell equations
for the coupling/probe envelope Rabi frequencies expressed in terms of the
pulse-localized coordinates $z$ and $\tau=t-z/c$,  $\Omega_{\rm
p}(z,\tau)$ and $\Omega_{\rm c}(z,\tau)$ reduce to \cite{eberly}
\begin{equation}
{\partial \over \partial z}\Omega_{\alpha} (z,\tau) = i \kappa_{ \alpha}
\tilde{\rho}_{\rm e\alpha} (z,\tau)
\label{maxwell}
\end{equation}
with $\alpha =(c,p)$ and where the parameter $\kappa_{\alpha}$ is given by
\begin{equation}
\kappa_{\alpha} = {\omega_{\alpha} N {\cal D}_{\rm e\alpha}^2\over c
\epsilon_0 \hbar}
\end{equation}
$N$ being the atomic density and $\epsilon_0$ the vacuum susceptibility.
The Beer's absorption length for the probe field $z_{\rm p}$ in the absence of
coupling laser is
\begin{equation}
\zeta_{\rm p}=\frac{A}{\kappa_{\rm p}}.
\label{penetr}
\end{equation}
In the ideal case, the solution of OBE and of Eq. (\ref{maxwell}) leads to a
shape-invariant propagation of the probe pulse
described  through the  dependence $\Omega_{\text p}(t-\frac{z}{v_{\rm
g}})$, $v_{\rm g}$ being the probe pulse group velocity.

OBE for the density matrix and Eqs. (\ref{maxwell}) for the laser propagation
were solved numerically
for an initial Gaussian pulse on the  probe transition
\begin{equation}
\Omega_{\text p} (z=0,t) = \Omega_{\rm
op}\exp\left(-t^2\over 2 T^2\right).
\label{impulso}
\end{equation}
 In the simulation the coupling laser, with constant Rabi frequency
$\Omega_{\rm c}$, was
assumed switched on at earlier times $t \le 0$, in
order to realize a counterintuitive pulse sequence.
The atoms were initially prepared in the ground $|{\rm p}\rangle$
state. Owing to the initial condition and the counterintuitive pulse
sequence, the atoms were initially prepared in the noncoupled state.

Because of the time dependence of the Rabi frequencies in the
noncoupled state of Eq.
(\ref{nonabsor}),  the adiabatic coherent preparation requires a proper
choice of the
time-scales\cite{bergmann,fleischh99,fleischh96}. In STIRAP the conditions
for the atoms in the
initial $|p\rangle$ state to remain in the noncoupled state following the
adiabatic evolution
are\cite{bergmann}
\begin{equation}
\Omega_{\alpha} \ge A;  T \ge 1/\Omega_{\alpha},
\label{berg}
\end{equation}
with $\alpha=(c,p)$. The first inequality, valid for resonant lasers,
imposes a minimum
Rabi frequency.  Larger Rabi frequencies are required  for detunings
$\delta_{\text c}$ and $\delta_{\text p}$ different from
zero\cite{bergmann}. The
second inequality imposes a pulse width long  compared to the inverse of
the Rabi frequencies.
We have verified numerically the role of both conditions on the slow
light propagation. However it should be noted that the STIRAP
conditions are required for an adiabatic transfer of the atomic
population, while for slow light the atomic response is not important,
and the only request is on the probe pulse propagation. For instance in the
limit of a very small $\Omega_{\text p}$ Rabi frequency, a negligible
STIRAP transfer is realized, still a slow light propagation
may be produced. In Ref.\cite{fleischh96} where  the formstable pulse
propagation was linked to the  preservation of population in the noncoupled
state,
the condition for a small nonadiabatic coupling between noncoupled
$|{\rm NC}(t)\rangle$ and coupled $|{\rm C}(t)\rangle$ states was
expressed  through the  following Rabi frequency
coupling $\Omega_{-}$:
\begin{equation}
\Omega_{-}=\frac{\dot{\Omega}_{\rm c}\Omega_{\rm p}-\dot{\Omega}_{\rm
    p}\Omega_{\rm c}}{\Omega^2},
\end{equation}
where the dot denotes the time derivative. On the basis of this
Rabi coupling, the adiabaticity condition required for a formstable
probe pulse propagation was written \cite{fleischh96}
\begin{equation}
\Omega_{-} \ll\Omega
\label{omegam}
\end{equation}
This relation, and the second one of Eq. (\ref{berg}), should be
satisfied for the slow light propagation.

The  probe Rabi frequency transmitted after the
propagation through the cold gas medium was determined numerically,
with  the pulse delay
$\tau_{\text d}(z)$ at the given spatial position $z$
defined by
\begin{equation}
\tau_{\text d} (z) = {\int_{-\infty}^{+\infty}\tau|\Omega_{\rm
p}(z,\tau)|^2
 d\tau \over
\int_{-\infty}^{+\infty}|\Omega_{\rm p}(z,\tau)|^2d\tau }.
\end{equation}
Finally the group velocity was calculated\cite{note}
\begin{equation}
v_{\rm g}=\frac{c}{1+\frac{c\tau_{\rm d}(z)}{z}}.
\end{equation}

The bandwidth $\Delta \omega$ of the EIT transparency  window
determines the spectral components of the probe pulse
propagating deep into the atomic medium without
shape distortion. The EIT window was explicitly investigated in the
applications of slow light to quantum entanglement  and stopping of
light \cite{lukin97}.
The transparency window for both closed and open systems depends on the decay
of the noncoupled state. For a closed system with $\gamma_{\rm  cp}=0$
at $\delta_{\text R}=0$ the transparency window is given
by
\begin{equation}
\Delta \omega=\frac{\Omega^2}{A}+\frac{1}{\Theta},
\end{equation}
whence the window decreases towards a limiting value at large interaction
times $\Theta$. For an open system that window is \cite{renzoni00}
\begin{equation}
\Delta \omega=\frac{\Omega^2}{2\sqrt{A\Theta}}
\sqrt{\left(1+\frac{\Gamma_{\rm p}}{\Gamma_{\rm out}}\right)
\frac{1}{\Omega_{\rm p}^2}
+\left(1+\frac{\Gamma_{\rm c}}{\Gamma_{\rm
out}}\right)\frac{1}{\Omega_{\rm c}^2}}.
\label{openwindow}
\end{equation}

\begin{figure}
\centering\begin{center}\mbox{\epsfxsize 3.2 in
\epsfbox{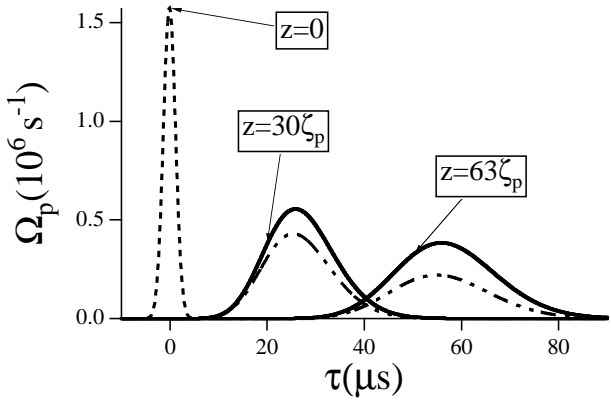}}
\end{center}
\caption{Probe Rabi frequency  $\Omega_{\rm p}(z,\tau)$ as a function
of the pulse-localized
time $\tau$ at different penetration distances  in  the conditions of
ref. \protect\cite{lau99}, with  $\Gamma_{\rm c}=A/3$,
$\Gamma_{\rm p}=A/2$, and $\Gamma_{\rm out}= A/6$.  The dashed line
represents a reference probe pulse with no atoms in the propagation
medium, scaled down by a factor two.
 The  continuous  and dashed-dotted lines are pulses for
propagation through the medium  at $z=30\zeta_{\text p}$ and
at $z=63\zeta_{\text p}$. Continuous lines for $\gamma_{\text cp}=0$ and
dashed-dotted lines for $\gamma_{\text cp}=10^{4}$ s$^{-1}$.
Initial Gaussian probe pulse  width  $T =80/A$, and  initial  coupling
laser Rabi frequency
$\Omega_{\text c}= 0.18A$.}
\label{pulse}
\end{figure}

For the  probe pulse propagation, where the interaction time $\Theta$
can be approximated by the pulse duration $T$, in order to have a
formstable pulse the transmission
window should be larger than the Fourier frequency distribution of the
pulse
\begin{equation}
\Delta \omega \gg 1/T.
\label{fourier}
\end{equation}
For an open system with bandwidth given by Eq. (\ref{openwindow}),
 Eq. (\ref{fourier}) imposes limitations on the Rabi frequencies.
That condition is satisfied by the parameters of typical slow light pulses
\cite{lau99,inouye00,liu01,scully99,budker99,phillips01}. Only for very
short pulses,
or very weak coupling lasers,  the EIT bandwidth
could produce a large distortion of the pulse shape propagating through a
three-level open system. We will derive in the next Section that,
owing to Eqs. (\ref{berg}) and (\ref{omegam}),  a
better transmission of the slow light pulses is obtained by increasing the
pulse duration $T$.

\section{Probe transmission}
\begin{figure}
\centering\begin{center}\mbox{\epsfxsize 3.2 in
\epsfbox{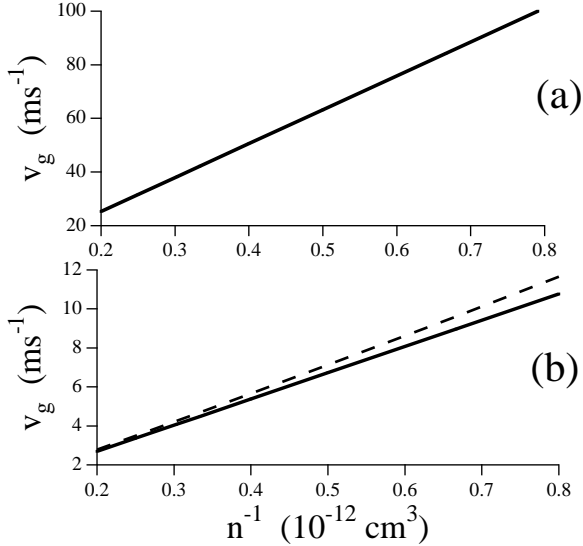}}
\end{center}
\caption{Probe-pulse velocity versus the inverse of the atomic
density. The data are in (a) for  coupling intensity of 12 mW/cm$^2$
corresponding to   coupling Rabi frequency $\Omega_{\text c}$ =0.56
$A$, and in (b) for a coupling
intensity of 3 mW/cm$^2$
corresponding to a  coupling Rabi frequency $\Omega_{\text c}$ =0.18
$A$.  The
continuous line represents results   for a simulation not including the atomic
momentum, while results of the dashed  line include in the simulation the
modifications
of atomic momentum produced by the light forces. In (a) continuous and
dashed lines cannot be
distinguished on the scale of the plot.}
\label{velocity}
\end{figure}

Numerical results for the propagation of the probe pulse within a
sample of open three level atoms are shown in Fig. \ref{pulse}. We used
the parameters of the  experiment in Ref.\cite{lau99}, {\it i.e.}
Na D$_2$ line parameters
($A = 2\pi\cdot 5.9$ MHz, $\lambda = 589.0$
nm),
cold atom density of $3.3\cdot10^{12}$ atoms/cm$^3$, propagation length
through the medium up to a distance $z$
equal to 63 times the Beer's length $\zeta_{\rm p}$, initial Rabi
frequencies $\Omega_{\rm
op}$ and $\Omega_{\rm c}$ around a tenth  of $A$.
The numerical results show a pulse peak height decreasing with the
penetration distance,  and  a probe pulse propagating with a  slow velocity.
We calculated from the data of Fig. \ref{pulse} the  slow
velocities shown in Fig. \ref{velocity},
i.e. in  the range of the values measured in ref.\cite{lau99}. In order to
compare these results
{\it quantitatively} with the response of a closed system, the atomic
parameters for
the closed system should be chosen carefully.
We may introduce a correspondence between closed and open systems by means
of a repumping laser.
As discussed previously,  we compared the slow light
propagation in an open system characterized by the $\rho_{\rm ee}$
decay rate $A$ of Eq. (\ref{open}), to that  of a closed system
characterized by the
decay rate $\Gamma^{'}_{\rm e}$ of Eq.  (\ref{opendecay}), the same optical
coherence decay
rate $A/2$ applying  to both systems. Numerical analyses for a closed
system corresponding to the open
system of Fig. \ref{pulse} have shown that  the same pulse transmission
and same slow light velocity  are obtained
for the two systems. In fact the pulse propagation of Fig.
\ref{pulse} was examined
for a counterintuitive coupling/probe pulse sequence and with laser
parameters producing a very small excited state occupation,
whence a small role of the $\Gamma_{\rm out}$ loss process.

We have examined the probe transmission modification produced by the
decoherence rate $\gamma_{\rm cp}$. The
decoherence process decreases the $|{\text  NC}(t)\rangle$ occupation, whence
decreases the probe transmission. We have verified that for
$\gamma_{\rm cp}$ larger than $0.001A$,
the probe transmission is well described through the EIT
 absorption length $\zeta_{\rm p}^{\rm EIT}$  given by the
following formula\cite{harris92}:
\begin{equation}
\zeta_{\rm p}^{\rm EIT} =
\frac{A}{\kappa_{\rm p}}
\left[\frac{\Omega_{\rm c}^2}{2\gamma_{\rm cp}A}+1\right]
\label{eitdepth}
\end{equation}

\begin{figure}
\centering\begin{center}\mbox{\epsfxsize 3.2 in
\epsfbox{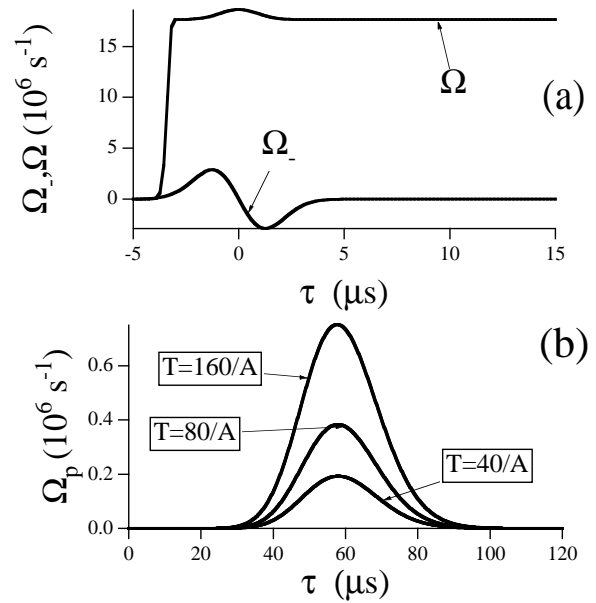}}
\end{center}
\caption{In (a) non-adiabaticity test comparison between $\Omega_{-}$
and $\Omega$ for $T =80/A$ at $z =0$. In (b) probe Rabi frequency
$\Omega_{\text p}$ as a function
of the pulse-retardation time $t$ at $z=63\zeta_{\text p}$  in the open
system of Fig. \ref{pulse} with different temporal widths of the probe
pulse. Higher transmission pulses are obtained  at $T =160/A$
decreasing transmission  at $T =80/A$ and $T =40/A$.}
\label{adiab}
\end{figure}

Note that  Eq. (\ref{eitdepth}) predicts no attenuation for the case
of $\gamma_{\rm cp}=0$. On the contrary the numerical results of Fig.
\ref{pulse} evidence
an attenuation of the probe pulse even in the
ideal case of zero ground state decoherence. In effect  for the
parameters of Fig. \ref{pulse} the adiabaticity condition of Eq.
(\ref{omegam}) required
for the complete validity of the EIT description is not fully satisfied.
The adiabaticity condition  was calculated  for
the laser pulses of Fig.  \ref{pulse}, as presented in  Fig. \ref{adiab}(a).
 $\Omega_-$ strongly depends on the value of
$T$, the probe pulse duration.  Better adiabatic
conditions, whence a larger probe pulse transmission, are realized
at larger $T$ values,  with longer duration pulses, as shown in Fig.
\ref{adiab}(b).
We verified that the probe-light slow velocity has a
weak dependence on the pulse length $T$, as in \cite{renzoni00}.  That weak
dependence
confirms that the dispersion slope of $Re(\rho_{\rm ep})$ versus the
laser detuning reaches a constant value at large
interaction time.

We have examined the dependence of the slow light velocity on the atomic
density $n$, in order to simulate the dependence on the cold atom
temperature investigated in \cite{lau99}. In the experimental investigation
the atomic density $n$ was varied with the sample temperature. Our results
for the slow
light velocity  are shown in Fig. \ref{velocity} for two different
values of the coupling laser parameters $\Omega_{\text c}$. A linear
dependence of the
velocity on the inverse of the atomic density is derived from the data,
with values between few m/s and 100 m/s depending on the atomic
density and the coupling laser Rabi frequency. The lowest velocities are
reached by decreasing the coupling laser Rabi frequency
$\Omega_{\text c}$.  For the dense media required  for
cold atom slow light propagation, the reabsorption of the spontaneously
emitted
photons, neglected in our theoretical analysis, may play an important
role. That reabsorption  is greatly reduced in the cigar shape
geometries of cold atomic samples\cite{ketterle}.

The numerical simulation allowed us to verify the presence of
adiabatons on the coupling laser, {\it i.e.} modifications in the
coupling laser intensity propagating through the atomic sample
synchronously with the probe laser \cite{adiabatons}. These modifications
are small, less than a ten percent of the coupling laser intensity.
Their observation in experiments as those of Refs.
\cite{lau99,inouye00,liu01,scully99,budker99,phillips01} would
represent a  confirmation of the coupling/probe matched pulse propagation.

\begin{figure}
\centering\begin{center}\mbox{\epsfxsize 3.2 in
\epsfbox{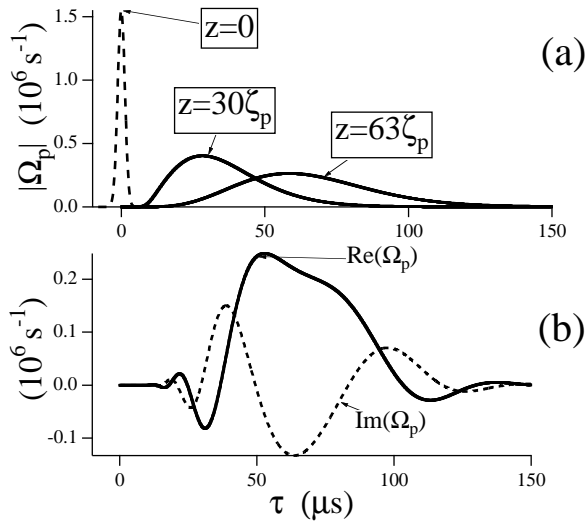}}
\end{center}
\caption{Propagation of a probe pulse slow light for coupling/probe lasers
detuned
from the upper $|e\rangle$ level, maintaining the Raman resonance
condition $\delta_{\text R}=0$, i.e. $\delta_{\text c}=\delta_{\text p} =
-A$, and  other parameters as in Fig. \ref{pulse}. In (a) absolute
value of the probe Rabi frequency, $|\Omega_{\text p}|$, versus the
pulse-localized time
$\tau$. As in Fig. \ref{pulse} the dashed line indicates a reference pulse.
Continuous lines for different propagation distances through
the medium. In (b) in-phase and out-phase components of the probe Rabi
frequency, $Re(\Omega_{\text p})$ and $Im(\Omega_{\text p})$ respectively,
at
propagation distance $z=63\zeta_{\rm p}$.}
\label{detuning}
\end{figure}

The probe pulse propagation was examined also for the case of laser
detunings $\delta_{\text c}$, $\delta_{\text p}$, different from zero,
satisfying the Raman resonance condition $\delta_{\text R}=0$. The
results of such simulation are shown in Fig. \ref{detuning}. For
laser detunings different from zero, the probe field Rabi frequency
acquires components in phase and out of phase with the initial probe laser
field,
and both components, $Re(\Omega_{\text p})$ and $Im(\Omega_{\text p})$
contribute to the pulse propagation. The absolute value
$|\Omega_{\text p}|$ is plotted in Fig. \ref{detuning}(a) while the
separate in-phase and out-phase components are shown in Fig. \ref{detuning}(b).
The individual in-phase and out-phase components
acquire complicate pulse shapes, even if the absolute Rabi frequency
propagates preserving the initial shape, apart from an attenuation
and a broadening.   The presence of a laser detuning
imposes stronger adiabaticity conditions, not well satisfied by the
weak Rabi coupling frequencies required for slow light propagation.
In fact,  comparing the results of Fig.
\ref{detuning}(a) with those of Fig. \ref{pulse}, it appears that even
for a small detuning the pulse shape of a slow pulse is
less preserved.    Increasing the laser detuning, the adiabaticity
condition can be satisfied increasing the coupling laser Rabi
frequency, at expenses of an increase of the slow velocity associated with
the probe
pulse.

\section{Atomic Momentum}
While the previous analysis neglected the atomic motion, in a sample
of cold atoms the modifications of the atomic
momentum produced by the laser interactions should be taken into
account.  The atomic states will be classified by the internal
variable $|\alpha\rangle$ with $(\alpha=c,p,e)$ and by the external
continuous variable $\vec{p}$ of the atomic momentum. For an atom initially
in the $|p\rangle$ state with momentum
$\vec{p}=0$, the Raman process of absorption and stimulated emission
transfers the atoms from state $|p,\vec{p}=0\rangle$ into the  state
$|c,\vec{k}_{\rm p}-\vec{k}_{\rm c}\rangle$. Including the atomic
momentum the noncoupled state at $t=0$ becomes
\begin{equation}
|{\rm NC}_{\rm m}(t=0)\rangle = {1\over \Omega}
\left( \Omega_{\rm c} |p,0\rangle
-\Omega_{\rm p} |c,\vec{k}_{\rm p}-\vec{k}_{\rm c}
\rangle \right).
\label{nonabsorp}
\end{equation}
The kinetic energy Hamiltonian operator $H_{\rm k}=p^2/2M$, $M$
being the atomic mass, couples the $|{\rm NC}_{\rm m}\rangle$
state to its orthogonal one $|{\rm C}_{\rm m}\rangle$ with a  rate
$\gamma_{\rm k}$ given by
\begin{equation}
\gamma_{\rm k}=
\frac{\langle {\rm C}_{\text m}|H_{\rm k}|{\rm NC}_{\text m}\rangle}{\hbar}=
\frac{\hbar \left( \vec{k}_{\text c}-\vec{k}_{\text p}\right)^2}{2M}
\frac{\Omega_{\rm p}^2}{\Omega^2}
\label{gammak}
\end{equation}
This rate $\gamma_{\text k}$ decreases the ground state atomic
coherence, and whence modifies the slowing down process of the light pulse.
For $\vec{k}_{\text c} \sim \vec{k}_{\text p}$, a rate $\gamma_{\text k}$
equal zero is realized
in the coupling/probe copropagating laser configuration, and this
result applies also  for an initial atomic momentum $\vec{p}$ different
from zero, as in the experiments of Refs. \cite{scully99,budker99}.
Instead the $\vec{k}_{\text c} \sim -\vec{k}_{\text p}$ counterpropagating
configuration was used in the laser cooling based on velocity-selective
coherent population
trapping  \cite{aspect}. For the experiment of \cite{lau99} with
orthogonal propagation directions for coupling and probe lasers, and
the parameters of Fig. \ref{pulse}, the
decoherence rate $\gamma_{\text k}$ is equal to 10$^{4}$ s$^{-1}$,
close to the value used in the
numerical analyses of Fig. \ref{pulse} with $\gamma_{\text c} \ne 0$.  It
should
be noted that the recoil frequency for sodium is $\omega_{\text
R}=\hbar k^2/2M = 1.7 \times 10^5$ s$^{-1}$, but in Eq.
(\ref{gammak}) the
$(\Omega_{\text p}/\Omega)^2$ factor produces a smaller $\gamma_{\text k}$.

In order to test the influence of the atomic momentum decoherence on
the slow light production, we solved numerically the generalized OBE,
{\it i.e.} the density matrix equations including the atomic momentum
\cite{aspect}, for the parameters of
the experiment in \cite{lau99}, i.e. with orthogonal $\vec{k}_{\rm c}$
and $\vec{k}_{\rm p}$. The numerical results for the slow
light velocity including the atomic momentum in the propagation are
shown in Fig. \ref{velocity}. The linear dependence
of $v_{\rm g}$ on $1/n$ remains also when the atomic momentum is
included in the analysis. The modification of the slow light velocity
produced by the atomic momentum is around ten percent at
the smaller coupling laser Rabi frequency.

\begin{figure}
\centering\begin{center}\mbox{\epsfxsize 3.2 in
\epsfbox{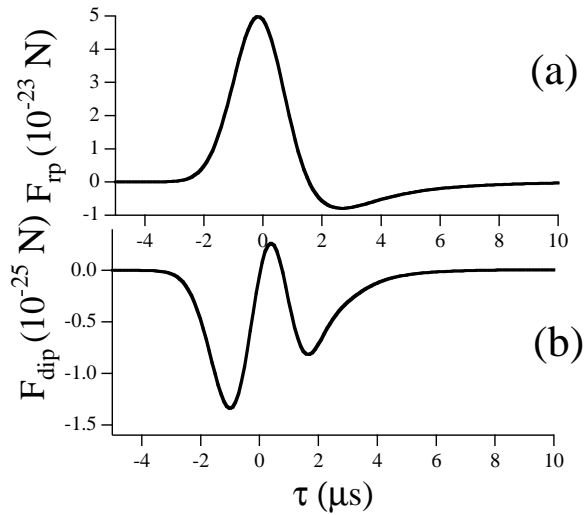}}
\end{center}
\caption{Forces acting on sodium cold atoms versus the pulse-localized
time $\tau$, at $\delta_{\rm c}=\delta_{\rm p}=-A$  while the probe
laser pulse of Fig. \ref{detuning} enters the propagating medium
(z=0).
Radiation-pressure force $F_{\rm rp}$ in (a), and dipole force
$F_{\rm dip}$ in (b).}
\label{forces}
\end{figure}

The modifications on the atomic momentum produced
by the laser pulse interaction originate from the forces acting
on the atoms. The longitudinal force, along the laser propagation
direction, on the atom of the slow light medium contains
both a radiation-pressure dissipative term and a dipole gradient
reactive
term\cite{cohen}. The dissipative  radiation-pressure force $F_{\rm
rp}$ acts on the atoms for both resonant and non resonant probe
lasers,  while the
reactive  dipole force $F_{\rm dip}$ is present only for
$\delta_{\text p} \ne 0$.  These forces are determined by
the optical coherence of the atomic dipole matrix \cite{three}:
\begin{mathletters}
\begin{eqnarray}
F_{\rm rp}(z,t) &=& -i\frac{\hbar}{2}k_{\rm p}\Omega_{\rm
p}(z,t)\rho_{\rm pe}^{*}(z,t)  + c.c. \nonumber \\
F_{\rm dip}(z,t) &=& \frac{\hbar}{2}\frac{\delta\Omega_{\rm p}}{\delta
z}\rho_{\rm
pe}^{*}  + c.c
\end{eqnarray}
\end{mathletters}
where, owing to the counterintuitive pulse sequence, and the
orthogonal $\vec{k}_{\rm c}$, $\vec{k}_{\rm p}$ geometry of
Ref. \cite{lau99}, only the probe laser contributes to the force along
the $z$ axis. These  forces, calculated for the propagating pulses of the
previous
figures, are  plotted versus time at $z=0$ in Fig. \ref{forces} for the
case of $\delta_{\rm c}=\delta_{\rm p}=-A$. From the
figure it appears that for the chosen laser detuning the radiation pressure
force is larger than
the dipole force by few orders of magnitude. Similar values were
obtained for the radiation pressure force for a probe laser in
resonance, when the dipole force is equal zero. From the point of
view of the atomic response, the total modification of the atomic momentum
depends on the integral of the force over the interaction time, i.e.
the pulse duration time. Because the radiation pressure force has an
antisymmetric dependence on the time, its integral, whence the
modification of the atomic momentum, is small. We have verified that
for same laser parameters, detunings and Rabi frequencies, for
instance those of Ref. \cite{harris00},  the time dependencies of the
light forces  are completely antisymmetric, so that the
forces produce a total modification of the atomic momentum equal zero.

\section{conclusions}
We have examined the propagation of a probe pulse through a medium composed
of an open three-level $\Lambda$ system in the presence of a coupling laser
acting on the adjacent transition of the system. Our numerical analysis
confirms
that also in an open system a large reduction of the probe-light
velocity is obtained.
The comparison with a closed system created by applying a
repumping laser to the open system has demonstrated that for the
parameters of the numerical analysis the velocities  of
the propagating pulse are equal in the compared open and closed
systems. Also the amplitudes of the transmitted probe pulses are similar
for the open and closed systems. This result corresponded to a set of
laser parameters where the excited state occupation was very small, so that
the external losses were not playing an important role on the atomic
evolution. The intensity of the coupling and probe lasers were
chosen in order to satisfy the requests of both EIT and excited state
occupation. Furthermore a proper choice of the width of the probe pulse
allowed us to satisfy the adiabaticity conditions of the atomic response,
with a larger pulse transmission.
We have examined the role of the laser pulse forces acting on the atoms. Those
forces modify the pulse propagation and for the
slowest pulses give an important contribution to the slow light velocity.
The contribution of the atomic momentum to the slow light propagation
can be tested using different laser propagation geometries, whence
different atomic momenta for the noncoupled state of the atomic
preparation.

\section{Acknowledgments}

E.A. acknowledges stimulating discussions with M. Fleischhauer on the
relations between STIRAP and EIT,
and with W.D. Phillips on the role of the atomic momentum.


\begin{references}
\bibitem{harris_rev} For a review S.E.~Harris, Physics Today {\bf 50},
36-42 (1997).
\bibitem{scullybook}See
also M.O.~Scully and M.S.~Zubairy, {\it Quantum Optics}, (University
Press, Cambridge, 1997).

\bibitem{arirev} For a review see E.~Arimondo, in Progress in Optics
ed. E. Wolf, vol. 35 (Elsevier, Amsterdam, 1996) p. 257-354.

\bibitem{lau99}L. V. Hau, S. E. Harris, Z. Dutton, and C. H.
Behroozi, Nature (London) {\bf 397}, 594-8 (1999).

\bibitem{inouye00} S.~Inouye, R.F.~Low, S.~Gupta, T.~Pfau,
A.~G\"orlitz, T.L.~Gustavson, D.E.~Pritchard, and W.~Ketterle, Phys.
Rev. Lett. {\bf 85}, 4225-8 (2000).

\bibitem{liu01}C.~Liu, Z. Dutton, C. H. Behroozi, and  L. V. Hau,
Nature (London) {\bf 409}, 490-3 (2001).

\bibitem{scully99}M.M.~Kash, V.A.~Sautenkov, A.S.~Zibrov, L.~Hollberg,
G.R.~Welch, M.D.~Lukin, Y.~Rostovtsev, E.S.~Fry, and M.O.~Scully,
Phys. Rev. Lett. {\bf 82}, 5229-5232 (1999).

\bibitem{budker99} D.~Budker, D.F.~Kimball, S.M.~Rochester, and
V.V. Yashchuk, Phys. Rev. Lett. {\bf 83}, 1767-1770 (1999).

\bibitem{phillips01} D.F.~Phillips, A.~Fleischhauer, A.~Mair,
R.L.~Walsworth, and M. D. Lukin, Phys. Rev. Lett. {\bf 86}, 783-6
(2001).

\bibitem{renzoni1} F.~Renzoni, W.~Maichen, L.~Windholz, and E.~Arimondo,
Phys. Rev. A {\bf55}, 3710-8 (1997).

\bibitem{morigi} The  slow light propagation within a condensate was
investigated by G.~Morigi and G.~Agarwal, Phys. Rev. A {\bf
62}, 013801 (2000).

\bibitem{bergmann} K.~Bergmann, H.~Theuer, and B.W.~Shore, Rev. Mod.
Phys. {\bf 70}, 1003-1025 (1998).

\bibitem{fleischh99}M.~Fleischhauer, Opt. Expr. {\bf
4}, 107-112 (1999).

\bibitem{fleischh96}M.~Fleischhauer and A.S.~Manka, Phys. Rev. A {\bf
54}, 794-803 (1996).

\bibitem{harris00} S.E.~Harris, Phys. Rev. Lett. {\bf 85}, 4032-5 (2000).

\bibitem{renzoni00} F.~Renzoni and E.~Arimondo, Opt. Commun. {\bf 178},
345-353  (2000).

\bibitem{renzoni98} F.~Renzoni and E.~Arimondo, Phys. Rev. A {\bf 58},
4717-4722 (1998).

\bibitem{eberly} M.~Sargent III, M.O.~Scully, and W.E.~Lamb, Jr, {\it
Laser Physics}, (Addison-Wesley, Reading, 1974); J.H.~Eberly, M.L.~Pons,
and H.R.~Haq, Phys. Rev. Lett. {\bf 72}, 56-9 (1994).

\bibitem{note} The velocity here defined is the centrovelocity
reducing to the group velocity for a quasi-monochromatic pulse. The
centrovelocity is the quantity measured in most experiments on slow
light.

\bibitem{lukin97} M.D.~Lukin, Phys. Rev. Lett. {\bf 79}, 2959-2962 (1997);
 M.D.~Lukin  and A.~Imamoglu, Phys. Rev. Lett. {\bf
84}, 1419-1422 (2000); O.~Kocharovskaya, Y.~Rostovtsev, and M.O.~Scully,
Phys. Rev. Lett. {\bf 85}, 628 (2001).



\bibitem{harris92} S.E.~Harris, J.E.~Field, and A.~Kasapi, Phys. Rev.
A {\bf 46}, R29-32 (1992).

\bibitem{ketterle} See recent review  by W. Ketterle, D.S.~Durfee, and
D.M.~Stamper-Kurn, in {\it Bose-Einstein condensation in atomic gases},
edited by M.
Inguscio, S.~Stringari and C.~Wieman (IOS Press, Amsterdam, 1999) p.
67-176.

\bibitem{adiabatons}R.~Grobe, F.T.~Hioe, and J.H.~Eberly Phys.
Rev. Lett. {\bf 73}, 3183-6 (1994).

\bibitem{aspect} A.~Aspect, E.~Arimondo, R.~Kaiser, N.~Vansteenkiste,
and C.~Cohen-Tannoudji, J.Opt. Soc. Am. B {\bf 6}, 2112-2124 (1989).

\bibitem{cohen} C.~Cohen-Tannoudji, in {\it Fundamental Systems in
Quantum Optics}, Proceedings of the Les Houches Summer School of
Theoretical Physics, Session LIII, edited by J.~Dalibard,
J.M.~Raimond, and J.~Zinn-Justin (Elsevier, Amsterdam, 1991) p. 1-172 .

\bibitem{three} The radiation pressure force and the dipole force for
a three-level $\Lambda$ system were examined by
P.V.~Panat and S.V.~Lawande, Phys. Rev. A {\bf 61}, 063406 (2000).

\end{references}
\end{document}